\documentclass{article}

\usepackage{arxiv}

\usepackage[utf8]{inputenc} 
\usepackage[T1]{fontenc}    
\usepackage[colorlinks=true]{hyperref}       
\usepackage{url}            
\usepackage{booktabs}       
\usepackage{amsfonts}       
\usepackage{nicefrac}       
\usepackage{microtype}      
\usepackage{graphicx}
\usepackage{amsmath}
\usepackage{amssymb}
\usepackage{comment}
\usepackage{breqn}
\usepackage{mathtools}

\graphicspath{{Figures/}}

\newcommand{\myspace}{\hspace{4pt}}

\title{Teaching GANs to Sketch in Vector Format}

\author{
  Varshaneya V \\
  Department of Mathematics and Computer Science\\
  Sri Sathya Sai Institute of Higher Learning\\
  Prasanthi Nilayam, Andhra Pradesh, India\\
  \texttt{varshaneya.v@gmail.com} \\
  \And
  S Balasubramanian \\
  Department of Mathematics and Computer Science\\
  Sri Sathya Sai Institute of Higher Learning\\
  Prasanthi Nilayam, Andhra Pradesh, India\\
  \texttt{sbalasubramanian@sssihl.edu.in}\\
  \AND
  Vineeth N Balasubramanian\\
  Department of Computer Science and Engineering\\
  Indian Institute of Technology Hyderabad\\
  Telangana, India\\
  \texttt{vineethnb@iith.ac.in}\\
}

\begin{document}
\maketitle

\begin{abstract}
Sketching is more fundamental to human cognition than speech. Deep Neural Networks (DNNs) have achieved the state-of-the-art in speech-related tasks but have not made significant development in generating stroke-based sketches a.k.a sketches in vector format. Though there are Variational Auto Encoders (VAEs) for generating sketches in vector format, there is no Generative Adversarial Network (GAN) architecture for the same. In this paper, we propose a standalone GAN architecture SkeGAN and a VAE-GAN architecture VASkeGAN, for sketch generation in vector format. SkeGAN is a stochastic policy in Reinforcement Learning (RL), capable of generating both multidimensional continuous and discrete outputs. VASkeGAN hybridizes a VAE and a GAN, in order to couple the efficient representation of data by VAE with the powerful generating capabilities of a GAN, to produce visually appealing sketches. We also propose a new metric called the Ske-score which quantifies the quality of vector sketches. We have validated that SkeGAN and VASkeGAN generate visually appealing sketches by using Human Turing Test and Ske-score.
\end{abstract}


\section{Introduction} \label{intro}
\vspace{-4pt}
Some of the very popular generative models in Deep Learning (DL) are VAEs \cite{vae} and GANs \cite{gan}. VAEs tend to maximize the likelihood of the generated data coming from the actual distribution while assuming a Gaussian prior. Different VAE architectures have performed well in generating various types of images ranging from handwritten digits \cite{vae,draw} to faces \cite{vae} to house numbers \cite{draw} and to CIFAR images \cite{draw}. On the other hand, in a GAN, the generator and discriminator play a minmax game until they reach an equilibrium. At this point, the distribution of generator is close to that of the original data. GANs are known to have been used for simple image generation in \cite{dcgan} as well as to more sophisticated tasks such as style-transfer \cite{cyclegan,yoo2016pixel}, super-resolution \cite{ledig2017photo} image-to-image translation \cite{pix2pix,pix2pixHD} and removing image in-painting \cite{pathakCVPR16context}. All of the aforesaid works are limited to only raster images.

Vector graphics were introduced in computer displays in the 1960s. Since then vector graphics have been studied very intensely. These images do not degrade when transformations are applied and they require minimal amount of space to be stored and transferred. Most importantly, they can be rescaled infinitely. They are represented as curves and strokes. 

Sketching was the means of communication much before languages were developed. Hence, sketching becomes a fundamental skill of human cognition. Today, DNNs perform the state-of-art in language-related tasks but there only a handful of works that discuss sketch generation in vector format, let alone vector image generation in the wild. As of today, there are  \cite{sketchrnn,ha2015recurrent,chen2017sketch,zhong} based on VAE, which generate sketches in vector format. This proves that sketch generation in vector format is indeed a very challenging problem to tackle. In addition to this, there are no GAN architectures for sketch generation in vector format.  So, we focus on generating sketches in vector format. 

We consider a sketch to be a collection of strokes, wherein each stroke consists of 2-D continuous offsets and 3-D discrete pen-states. This representation is also known as the \textit{stroke-based format}. The discrete outputs pose a difficulty for the gradient updates to be passed from discriminator to generator for the weight update. \cite{seqgan} proposed a novel policy gradient based loss for generating 1-D discrete tokens. Whereas in our case, we have a combination of 2-D continuous variates and 3-D discrete variates which makes adapting of policy gradient loss not a straight-forward task. Currently there are no metrics that quantify the `goodness' of vector sketches. Hence, we have proposed the `Ske-score' which quantifies the goodness of vector sketches. Our contributions are as follows:
\vspace{-4pt}
\begin{itemize}
\setlength\itemsep{-1pt}
    \item The first GAN-based architecture \textbf{SkeGAN}, for sketch generation in vector format. To this end we propose a novel coupling mechanism which models the influence of offsets on pen-states while sketching.
    \item An alternative GAN architecture \textbf{VASkeGAN} based on VAE-GAN architecture \cite{vaegan} for comparison with SkeGAN.
    \item A new metric known as the \textbf{Ske-score}, which quantifies the goodness of generated vector sketches.
\end{itemize}

\section{Related Work}
\vspace{-4pt}
There are a very few approaches of sketch generation which use stochastic techniques such as Hidden Markov Models (HMMs) \cite{sketchInterpretationRefinement} and others that use pure image processing techniques such as \cite{robotdrawing}. There are quite a lot of work done relating to human-drawn sketches in general using DL such as recognition \cite{sketchrecognition,sarvadevabhatla2016analyzing,sarvadevabhatla2015eye}, eye-fixation or saliency \cite{eyefixation}, guessing a sketch being drawn \cite{pictionaryWord} and parsing \cite{sketchparse}. Specifically \cite{gao2017ca-gan} uses GANs to generate sketches of human faces given digital portraits of their faces. There are also works such as \cite{chen2018sketchygan,lu2018image} which discuss the approaches to convert rasterized sketches into realistic images. One commonality amongst all of them is that all of them work with sketches in the raster format.

The first attempt in generating vector images is by \cite{ha2015recurrent} to generate Kanji (Chinese alphabet) characters using a two-layered LSTM, where each Kanji character is represented in the stroke-based format. Following \cite{ha2015recurrent}, D. Ha et al. propose a VAE model called the ``sketch-rnn", for vector sketch generation in \cite{sketchrnn}, which is trained on the ``QuickDraw" dataset \cite{quickdrawdataset}. Here too, the sketches are represented in the stroke-based format. This paper shows very good performance in unconditional generation, conditional reconstruction, latent space representation and predicting the ending of incomplete sketches for a variety of classes of objects. \cite{sketchrnn} produced visually appealing sketches when trained with a single category of sketch. The sketches are not visually appealing when a mix of category is used for training. Hence, in order to overcome this difficulty, \cite{chen2017sketch} replaces the encoder of \cite{sketchrnn} with a Convolution Neural Network(CNN) and removes the KL-Divergence loss. This model too produces sketches in the stroke-based format. Since the convolution is spatial, the input to the this model is rasterized format of sketches from the QuickDraw dataset. Based on the Turing Test, the authors of \cite{chen2017sketch} conclude that the models with CNN encoders outperformed those with RNN encoders in generating human-style sketches. K. Zhong in \cite{zhong} extends the VAE proposed in \cite{sketchrnn} to create an end-to-end pipeline which takes in fonts in Scalable Vector Graphics (SVGs) to learn and generate novel fonts. The results are demonstrated on Google Fonts Dataset.

\begin{figure}[]
\centerline{\includegraphics[scale=.22]{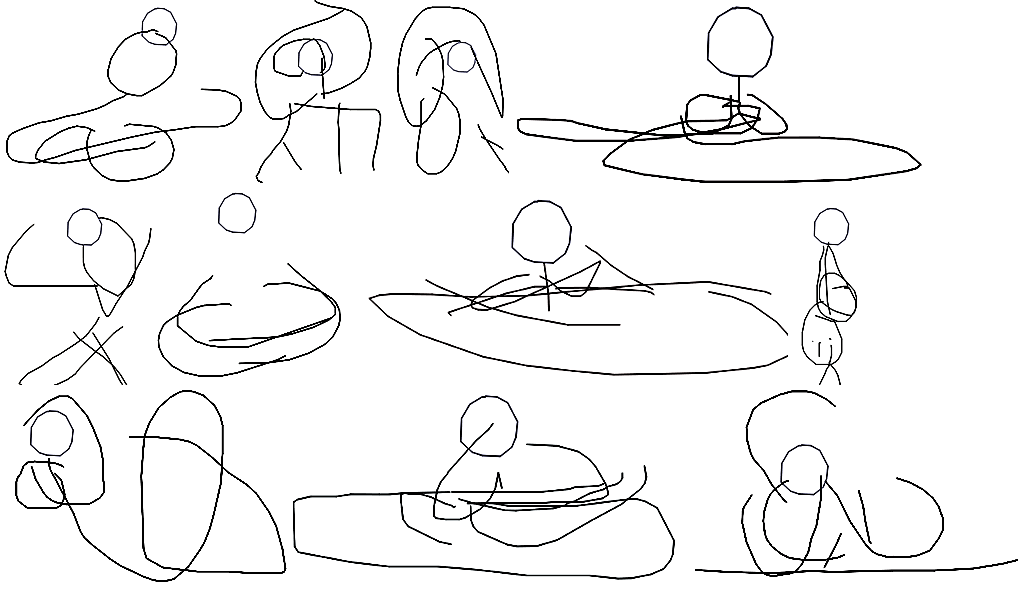}
            \includegraphics[scale=.22]{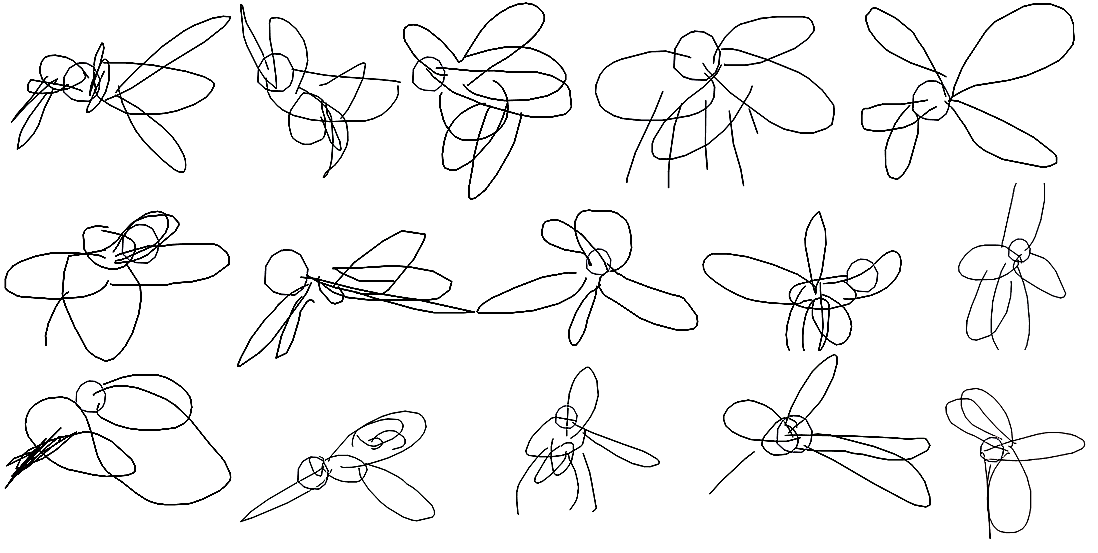}}
\caption{Scribble effect of \cite{sketchrnn} with yoga pose sketches (left) and mosquito sketches (right).}
\label{scribble}
\end{figure}

All the architectures mentioned for sketch generation in vector format, are VAEs. A well known disadvantage with VAEs, they tend to produce blurred images in case of raster images. Since there is no concept of blurring, the vector images produced by VAEs like sketch-rnn \cite{sketchrnn} tend to suffer from a mode-collapse-like situation wherein the pen is not lifted to draw at another location, but stays on the paper and continues to scribble. We call this as the \textit{scribble effect}. Figure \ref{scribble} shows this effect in the sketches of ``yoga poses" and ``mosquitos". Since VAEs assume the prior to be Gaussian, they need to be trained for a very large number of iterations so that the weights of the decoder get adjusted accordingly in order to generate close to the distribution of data. In the case of \cite{sketchrnn}, the training is done for 10 million iterations. Also, GANs have performed outstandingly well for a variety of tasks mentioned in Section \ref{intro}, with raster images. In order to alleviate these disadvantages of VAEs and harness the power of GANs, we propose a standalone GAN called the \textbf{SkeGAN} and another GAN called the \textbf{VASkeGAN} with which we compare SkeGAN.

\section{Our Contributions}
\vspace{-4pt}
\subsection{Ske-score: Evaluation Metric}
\vspace{-3pt}
Since sketches suffer from scribble effect, we propose a novel metric known as the `Ske-score' which quantifies the goodness of a sketch. Ske-score $\mathcal{S} \in \mathbf{R}^+_0$, of a sketch is defined as the ratio of number of times the pen is lifted to number of times it touches the paper while the sketch is being drawn. The Ske-score of a model $\mathcal{S}_M$ is defined to be the average of individual Ske-scores $\mathcal{S}$ of sketches generated by it. The Ske-score of the dataset $\mathcal{S}_D$ is the average of Ske-scores of all of the sketches in it. Intuitively, this metric quantifies the fraction of time when pen is lifted from the paper. A high value of $\mathcal{S}$ indicates that pen is lifted more number of times. A model $M$ is said to generate `good sketches' without scribble effect iff $\mid \mathcal{S}_D - \mathcal{S}_M \mid < \epsilon$.

\subsection{Problem Setup}
\vspace{-3pt}
Sketches are considered to be a collection of 5-tuple $(\Delta x, \Delta y, q_1, q_2, q_3)$, where $(\Delta x, \Delta y)$ are the offsets to be moved along $X$ and $Y$ axes respectively and $(q_1, q_2, q_3)$ is the pen-state. $(q_1, q_2, q_3)=(1,0,0)$ indicates that the pen is on the paper, $(q_1, q_2, q_3)=(0,1,0)$ indicates it is lifted and $(q_1, q_2, q_3)=(0,0,1)$ indicates that the drawing has ended. Pen-state is modeled as categorical random variable. All the drawings are assumed to start from the origin. This is done by prepending the sketch with the start-of-sequence symbol $S_0$, which is $(0,0,1,0,0)$. The offsets are modeled as a Gaussian Mixture Model (GMM) in the case of SkeGAN and as IID normal variable in the case of VASkeGAN. In both the models, we incorporate the parameter $\tau \in [0,1]$ as defined in \cite{sketchrnn}, to control the randomness or the variety in the generated samples. Sketch generation is done tuple-by-tuple until $q_3$ is not 1 or until the maximum length $N_{max}$ is reached. A sketch is generated stroke by stroke, wherein the stroke at time-step $i$ depends on all of the strokes at previous time-steps. In order to model this dependency, the generator is an auto-regressive model like LSTM or GRU. The discriminator distinguishes whether a batch of sketches has come from the dataset or from the generator. So it must understand the dependency between strokes of different time-steps in order to distinguish the sketches. Therefore, the discriminator is also an auto-regressive model like LSTM or GRU.

\subsection{SkeGAN: A Sequential GAN for Vector Images }
\vspace{-4pt}
In a GAN architecture, the weights of the generator are updated based on the signal/reward from the discriminator. GANs have a limitation when there is a need to generate discrete tokens. The discrete outputs pose a difficulty for the gradient updates to be passed from discriminator to generator for the weight update. In our case, the offsets are continuous random variables whereas the pen-states are discrete random variables. So, given a conventional GAN architecture, during the back-propagation the gradient updates are passed without any difficulty for the offsets but not for pen-states. Also, any discriminator can guide a generator only when a complete sequence is given to it. This means that the discriminator cannot guide the generator while it is in the process of generating a sequence. Also, in our case we must generate both discrete and continuous data. Therefore, we propose a coupled GAN architecture with a combination of policy gradient and standard adversarial losses to generate both multi-dimension discrete and continuous tokens. The generator $G$ in SkeGAN is a stochastic policy in RL which can sample tuples for the Monte Carlo search. By performing a Monte Carlo search, the reward signal from discriminator $D$ is passed back to $G$ even at its intermediate action value. Further, policy gradients are used for updating the weights of $G$ via gradient ascent mentioned in Equation \ref{gradAscent}.

\begin{figure}[]
\centerline{\includegraphics[scale=0.27]{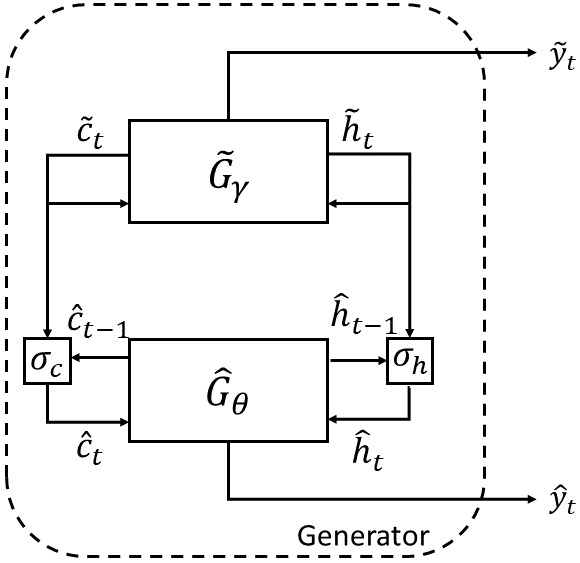}
\includegraphics[scale=0.27]{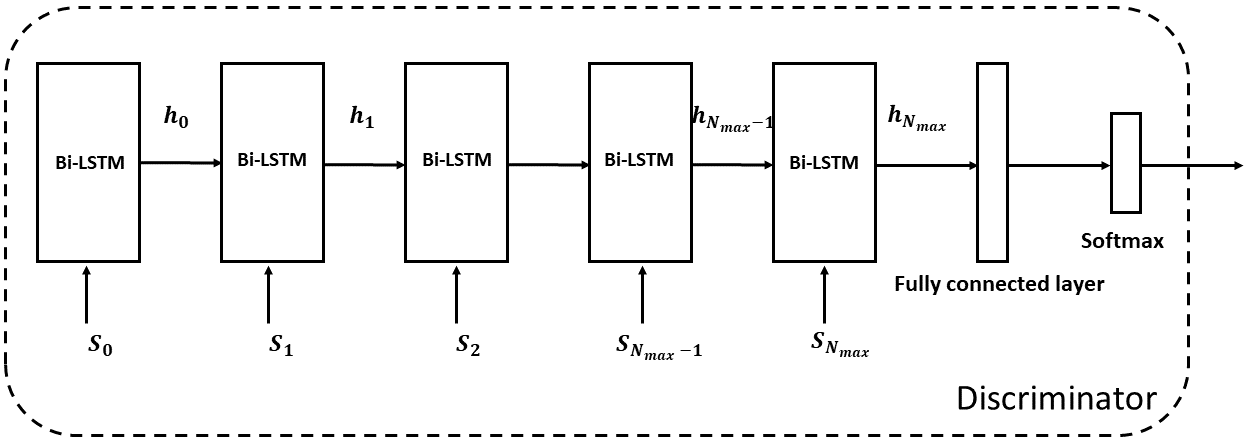}}
\caption{Generator (Left) and Discriminator (Right) of SkeGAN}
\label{seqganGenDis}
\end{figure}

We assume that the current coordinate at which the pen is situated dictates whether the pen must be on the paper or must be lifted, when it is to be moved to the next coordinate. In other words, we assume that the offsets influence the pen-states. In addition to this, the previous pen-state influences the next pen-state. So, the current pen-state depends both on its previous state and the current offset. To model this relationship, we propose a coupled generator $G$ consisting of two generators viz. $\tilde{G}$ for generating offsets and $\hat{G}$ for generating pen-states. Each of $\hat{G}$ and $\tilde{G}$ is an LSTM with a hidden size of 512. The hidden state of $\tilde{G}$ at time-step $t$ is denoted as $\tilde{h}_t$ and that of $\hat{G}$ at time-step $t$ is denoted as $\hat{h}_t$. The coupling is achieved by having two update gates $\sigma_c$ and $\sigma_h$ with learnable parameters. The coupling effect can be mathematically described in the following equations:
\begin{eqnarray}
\hat{h} & = & \sigma_h(W_h[\tilde{h}_t,\hat{h}_{t-1}]+b_h) \\
\hat{h}_t & = & \hat{h} \odot \tilde{h}_t + (1-\hat{h}) \odot \hat{h}_{t-1} \\
\hat{c} & = & \sigma_c(W_c[\tilde{c}_t,\hat{c}_{t-1}]+b_c) \\
\hat{c}_t & = & \hat{c} \odot \tilde{c}_t + (1-\hat{c}) \odot \hat{c}_{t-1}
\end{eqnarray}
where $\odot$ refers to element-wise multiplication and $W_h$, $W_c$, $b_h$ and $b_c$ are learnable parameters. The Generator of the proposed architecture is shown in the left portion of Figure \ref{seqganGenDis}. At each time-step $t$, $\tilde{G}$ generates $\tilde{y}_t$ and $\hat{G}$ generates $\hat{y}_t$. The parameters for the distribution of offsets are estimated from $\tilde{y}_t$ while those for the distribution of pen-states are estimated from $\hat{y}_t$ as given in \cite{sketchrnn}.

The discriminator $D$ is a Bidirectional LSTM with a hidden size of 256. A batch with one half containing generated sketches and another half from the dataset is shuffled and given to it. The forward and the backward hidden states of the LSTM are concatenated and mapped to a vector of dimension 2 followed by softmax activation to predict the probability of each sequence being real or fake. The discriminator is shown in the right portion of Figure \ref{seqganGenDis}.

\vspace{-2pt}
\paragraph*{Policy gradient based formulation:}

Let $G = (\tilde{G}_\gamma,\hat{G}_\theta)$ and $D = D_\phi$. Since this policy gradient based formulation is meaningful only for discrete tokens, the following discussion pertains to $\hat{G}_\theta$ alone. Given a sequence $Y = (y_1,y_2, ... ,y_T)$, where each $y_t$ is a 3-tuple consisting of a valid pen-state i.e. $y_t \in \mathcal{Y}$ and $\mathcal{Y}=\{(1,0,0),(0,1,0),(0,0,1)\}$. At a time-step $t$, the state $s$ of $\hat{G}_\theta$ is the sequence of produced tokens, which is $(y_1,y_2, ... ,y_{t-1})$ and its action $a$ is to select the next token $y_t$. It can be observed that though the policy model $\hat{G}_\theta(y_t \mid Y_{1:t-1})$ is stochastic, the transition state is deterministic after an action. In other words, if $s = Y_{1:t-1}$ is the current state and the action is $y_t$, then next state $s' = Y_{1:t}$. $D_\phi(1:Y_t)$ is the probability of a particular sequence being real or not. Since there is no intermediate reward for an incomplete sequence, the objective $\hat{G}_\theta$ is to generate a sequence from the start state $s_0$ which maximizes the expected end reward as given by:
\begin{equation}
J(\theta) = \mathbb{E}[R_T \mid s_0, \theta] = \sum_{y' \in \mathcal{Y}} \hat{G}_\theta(y' \mid s_0) \cdot Q_{D_\phi}^{\hat{G}_\theta} (s_0,y')
\end{equation}
where $Q_{D_\phi}^{\hat{G}_\theta} (s_0,y')$ is the action-value function of the sequence and $R_T$ is the reward of the complete sequence. $Q_{D_\phi}^{\hat{G}_\theta} (s_0,y')$ is the expected accumulative reward starting from state $s$, taking action $a$ by following the policy $\hat{G}_\theta$. The next step is to estimate the action-value function. The estimated probability of a sequence being real $D_\phi(Y^n_{1:T})$, is considered to be the reward for $\hat{G}_\theta$. The $D_\phi$ can provide the reward for a complete sequence only. Also, one must look for maximizing the long-term rewards. Therefore to evaluate every intermediary step $t$, Monte Carlo search with a rollout policy $\hat{G}_\beta$ is used to sample the rest of the $T-t$ tokens. Let the output of an $N$-time Monte Carlo search be represented as:
\vspace{-5pt}
\begin{equation}
MC^{\hat{G}_\beta} (Y_{1:t};N) = \big\{Y^1_{1:T}, ... , Y^N_{1:T}\big\}
\vspace{-3pt}
\end{equation}
where $Y_{1:t}^n = (y_1, ..., y_t)$ and $Y^n_{t+1:T}$ is sampled based on rollout policy and the current state. Here $\hat{G}_\beta$ is set to $\hat{G}_\theta$ itself for simplicity and speed. Thus, the action value function for $\hat{G}_\theta$ is defined as:
\vspace{-5pt}
\begin{dmath} \label{MCRollout} 
Q_{D_\phi}^{\hat{G}_\theta}(s = Y_{1:t-1},a = y_t) = 
\begin{cases*}
D_\phi(Y_{1:T}^n), \quad Y_{1:T}^n \in MC^{\hat{G}_\beta}(Y_{1:t};N) & \text{for } t $\leq$ T \\
D_\phi(Y_{1:t}), & \text{for } t = T
\end{cases*}
\end{dmath}

The advantage of using $D_\phi$ as the reward function is that, since it is updated by the adversarial loss at every iteration, it improves its capability of distinguishing between real and fake. Due to this, it can provide better feedback to $G_\theta$. The gradient of the loss function with respect to $\theta$ from \cite{sutton} is given by:

\begin{dmath}
\nabla_\theta J(\theta) = \sum^T_{t=1} \mathbb{E}_{Y_{1:t-1} \sim \hat{G}_\theta} \bigg[ \sum_{y_t \in \mathcal{Y}} \nabla_\theta  \hat{G}_\theta{(y_t \mid Y_{1:t-1})} \cdot Q_{D_\phi}^{\hat{G}_\theta} (Y_{1:t-1},y_t) \bigg]
\end{dmath}
Using likelihood ratios \cite{seqgan}, $\nabla_\theta J(\theta)$ becomes:
\vspace{-5pt}
\begin{dmath}
\nabla_\theta J(\theta) \simeq \sum^T_{t=1} \mathbb{E}_{y_t \sim \hat{G}_\theta(y_t \mid Y_{1:t-1}) } \big[ \nabla_\theta  \hat{G}_\theta{(y_t \mid Y_{1:t-1})} \cdot Q_{D_\phi}^{\hat{G}_\theta} (Y_{1:t-1},y_t) \big]
\end{dmath}
The parameters of $\hat{G}_\theta$ are updated by the gradient ascent rule which is also known as policy gradient equation:
\vspace{-4pt}
\begin{equation} \label{gradAscent}
\theta \leftarrow \theta + \alpha_h\nabla_\theta J(\theta)
\vspace{-3pt}
\end{equation}
where $\alpha_h \in \mathbb{R}^+$ is the learning rate at the $h^{th}$ iteration.

\vspace{-3pt}
\paragraph*{Training:}
In order to ensure stability and faster convergence, the training of SkeGAN is done in two stages viz. the pre-training and the adversarial training. $G$ is pre-trained so that it can avoid generating meaningless values for both offsets and pen-states. At each time-step $i$, $S_{i-1}$ is fed to $G$. The pre-training of $G$ is done for 38500 iterations and the loss function used is the reconstruction loss $L_R$ \cite{sketchrnn} as given in Equation \ref{lr}. $D$ is also pre-trained so that it can effectively differentiate between the real and the fake samples to provide better feedback to $G$. In our case, $D$ is pre-trained for 35000 iterations. Each batch for pre-training $D$ contains 50\% samples from the dataset (labeled as real data) and 50\% samples generated by $G$ (labeled as fake data). The loss function used is the binary cross entropy loss. The number of iterations to pre-train was decided using empirical studies.

The adversarial training is done as in \cite{seqgan}. One round of training constitutes one epoch (700 iterations) of training $G$, followed by two epochs (1400 iterations) of training $D$. At each iteration in the training of $G$, the policy gradients based loss is used to update the parameters of $\hat{G}_\theta$ and adversarial loss is used to update the parameters of $\tilde{G}_\gamma$. Firstly, a sequence $Y_{1:T} = (y_1, ... , y_T )$ is generated from $G$. The action value function $Q(a = y_t;s=Y_{1:t-1})$ is then calculated for $1 \leq t \leq T$ by using the Monte Carlo rollout Equation \ref{MCRollout}, only for the pen-states. The parameters of $\hat{G}_\theta$ are updated by policy gradient Equation \ref{gradAscent}. For finding the adversarial loss, a sequence $(S_0, ... ,S_{N_{max} - 1})$ of length $N_{max}$ is given to $G$. The generated sequence is then prepended with $S_0$ and a batch containing such sequences is given to $D$ for its decision. Based on this decision, the weights of $\tilde{G}_\gamma$ are updated. The training of $D$ is similar to its pre-training.

As in \cite{sketchrnn}, the offsets $(\Delta x,\Delta y)$ are modeled as a mixture of $M$ bivariate normal distributions with the parameters for each of the distribution $i$ to be $(\mu_{x,i},\mu_{y,i},\sigma_{x,i},\sigma_{y,i},\rho_{xy,i})$. There is an additional vector $\Pi$ of length $M$ which consists of the mixing weights for the Gaussian Mixture Model (GMM). Therefore, at each time-step $t$, hidden state of $\tilde{G}$ is mapped to a vector $\tilde{y}$ of size $6M$, from where the parameters of the GMM are sampled. The distribution of the offsets are given as:

\begin{equation}
p(\Delta x,\Delta y) = {\sum_{j=1}^M \Pi_j \mathcal{N}(\Delta x,\Delta y \mid \mu_{x,j},\mu_{y,j},\sigma_{x,j},\sigma_{y,j},\rho_{xy,j})}, \quad \text{where} \quad \sum_{j=1}^M \Pi_j = 1
\end{equation}

The vector $\tilde{y}$ can be split into the parameters of the GMM as:
\begin{equation}
[(\tilde{\Pi}_1,\mu_x,\mu_y,\tilde{\sigma}_x,\tilde{\sigma}_y,\tilde{\rho}_{xy})_1, ..., (\tilde{\Pi}_M,\mu_x,\mu_y,\tilde{\sigma}_x,\tilde{\sigma}_y,\tilde{\rho}_{xy})_M] = \tilde{y}
\end{equation}
The weight for each of the component $k$ in the GMM is calculated as:
\begin{equation}
\Pi_k = \frac{exp(\tilde{\Pi}_k)}{\sum_{j=1}^M exp(\tilde{\Pi}_k)}, \myspace k \in \{1,2, ..., M\}
\end{equation}
We then apply $exp$ and $tanh$ operations to ensure that the standard deviations are non-negative and correlation is in the range $[-1,1]$.
\begin{equation}
\sigma_x = exp(\tilde{\sigma}_x), \quad \sigma_y = exp(\tilde{\sigma}_y), \quad \rho_{xy} = tanh(\tilde{\rho}_{xy})
\end{equation}

The pen-states is a vector of size 3 and hence the hidden state of $\hat{G}$ is mapped to a vector $\hat{y}$ of size 3. The probabilities are calculated for pen-states as:

\begin{equation}
q_k = \frac{exp(\hat{q}_k)}{\sum_{j=1}^3 exp(\hat{q}_k)}, k \in \{1,2,3\}, \quad \text{where} \myspace (\hat{q}_1, \hat{q}_2,\hat{q}_3) = \hat{y}
\end{equation}

$\tilde{y}$ and $\hat{y}$ are concatenated to get $y \in \mathbb{R}^{6M+3}$. We incorporate the parameter $\tau \in [0,1]$ as in \cite{sketchrnn}, to control the randomness or the variety in the generated samples. Mathematically writing, the parameters $\hat{q}_k$, $\sigma_x^2$ and $\sigma_y^2$ would be replaced by $\frac{\hat{q}_k}{\tau}$, $\sigma_x^2\tau$ and $\sigma_y^2\tau$ respectively.

The number of mixtures $M$ in the GMM is 20. The batch size is set to 100. $N_{max}$ is set as the length of the longest sequence  in the dataset. Gradients are clipped between $[-1,1]$ for both $G$ and $D$ to avoid exploding of gradients, which is a common issue with sequence models. The initial learning rate is set to 0.001, with a decay of 0.9999 after every 700 iterations for $G$ and 1400 iterations for $D$. The learning rate is decayed only if it is above  0.00001. Recurrent dropouts with a drop probability of 0.1 is used. The maximum number of steps in the rollout is set to 8 and the update rate for the policy gradient update is set to 0.8. Parameters of $G$ are updated using Adam optimizer and those of $D$ are updated using Stochastic Gradient Descent (SGD).

\begin{figure*}[]
\centerline{\includegraphics[scale=.4]{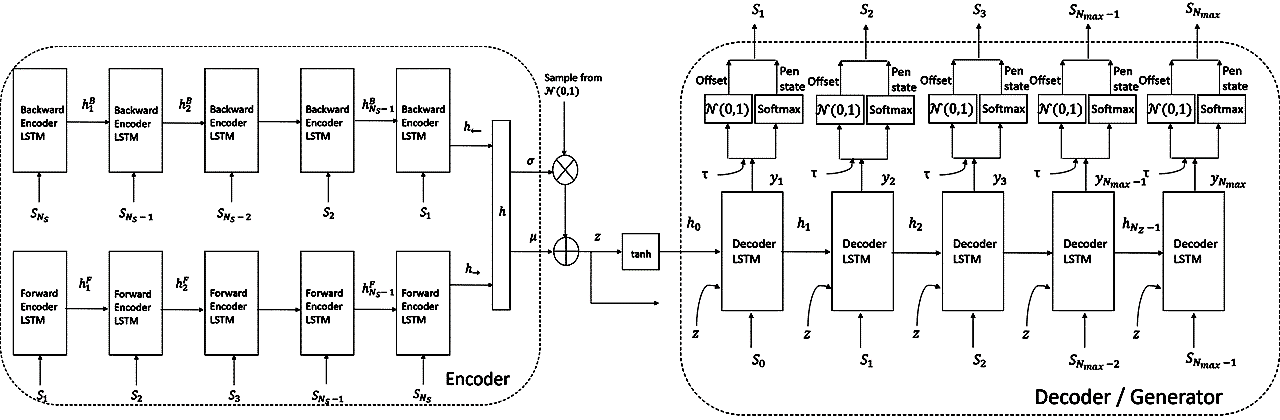}}
\caption{Encoder and Decoder of VASkeGAN architecture.}
\label{vaeganencdecarch}
\end{figure*}

\begin{figure}[]
\centerline{\includegraphics[scale=.2]{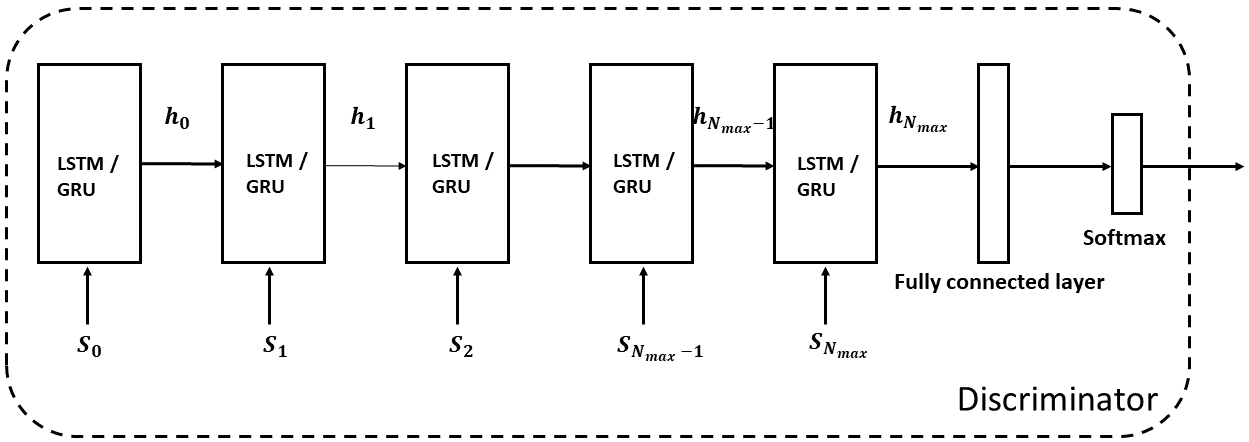}}
\caption{Discriminator of VASkeGAN architecture.}
\label{vaegandisarch}
\end{figure}

\subsection{VASkeGAN: VAE-GAN for Sketch Generation}
\vspace{-3pt}
For the sake of fair comparison of SkeGAN with another GAN architecture, we propose a VAE-GAN \cite{vaegan} based architecture. Since VAEs are good at representing the data in the latent space and GANs at generating data, the VAE in VAE-GAN produces meaningful representation of the data, which helps the generator in generating data close to its actual distribution. \cite{vaegan} shows the success of VAE-GAN architecture for datasets such as \textit{CelebA} \cite{celeba} and Labeled Faces in the Wild (LFW) \cite{LFW}. It alleviates the blurring in generated images. We hence propose \textit{VASkeGAN} for sketch generation.

VASkeGAN is a combination of VAE and GAN, wherein the decoder of the VAE doubles up to be the generator of the GAN and there is a discriminator. The encoder is a Bi-directional LSTM with a hidden size of 256, which takes a sketch as an input and produces a latent vector of size $N_z$. The parameters $\mu$ and $\hat{\sigma}$ are estimated from $h$ using linear layer with learnable parameters, where $h$ is a concatenation of the forward and backward hidden states. $\sigma$ is calculated from $\hat{\sigma}$ as $\sigma = exp(\frac{\hat{\sigma}}{2})$. $\mu$ and $\sigma$ are then used to obtain the latent vector  $z \in \mathbb{R}^{N_z}$ based on the ``reparametrization trick" given as $z = \mu + \sigma \odot \mathcal{N}(0,I)$. The decoder/generator is an LSTM with a hidden size of 512 and it produces sketches conditioned on $z$. The initial hidden state $h_0$ and cell state $c_0$ are derived from $z$ via the following equation: $[h_0;c_0] = tanh(W_z Z + b_z)$, where $W_z$ and $b_z$ are learnable parameters. The input $x_i$ to the decoder at each time-step $i$ is the previous point $S_{i-1}$ concatenated with $z$. The output of decoder $y_i \in \mathbb{R}^7$ at every time-step $i$, can be split as:
\begin{equation}
[ \mu_{x,i}, \mu_{y,i}, \hat{\sigma}_{x,i}, \hat{\sigma}_{y,i}, (\hat{q}_{1,i},\hat{q}_{2,i},\hat{q}_{3,i})] = y_i
\end{equation}
Exponential operation is applied to $\hat{\sigma}_{x,i}$ $\hat{\sigma}_{y,i}$ so that the standard deviations are non-negative and softmax is used to calculate the probabilities of the pen-states. The sampling of $\Delta x_i$ and $\Delta y_i$ is akin to the reparametrization trick. In other words, $\Delta \hat{x}_i \sim \mathcal{N}(0,1)$ and $\Delta \hat{y}_i \sim \mathcal{N}(0,1)$ are independently sampled and $\Delta x_i$ and $\Delta y_i$ are calculated as:
\begin{equation}
\Delta x_i = \Delta \hat{x}_i \cdot \sigma_{x,i} + \mu_{x,i} \quad \Delta y_i = \Delta \hat{y}_i \cdot \sigma_{y,i} + \mu_{y,i}
\vspace{-3pt}
\end{equation}

We experimented with two types of discriminators viz. one being a GRU with a hidden size of 512 and another being an LSTM with a hidden size of 512. The discriminator has to classify whether the batch of sketches presented to it is from the actual dataset (real samples with distribution $p_{data}$) or is generated by the generator (fake samples with distribution $p_\beta$). The hidden layer in the last time-step is mapped to a feature vector of size 2 using a linear layer with softmax activation, so as to output class probabilities of it being real and fake. The GAN plays a minmax game with the generator trying to ``fool" the discriminator, while the discriminator tries to foil this attempt. The model is fully trained when the discriminator is no longer able to detect whether the batch of samples is from the real data or generated by the generator. At that time, $p_\beta$ is closer to $p_{data}$. The proposed architecture is shown in Figures \ref{vaeganencdecarch} and \ref{vaegandisarch}.

\vspace{-3pt}
\paragraph*{Training}

The training of VAE-GAN is a combination of the training procedures of VAE and GAN. We train the encoder and the decoder (generator $G$), with reconstruction loss $L_R$ and KL Divergence loss $L_{KL}$ from \cite{sketchrnn} and the adversarial loss for generator $L_{adv}^G$. The adversarial loss for $G$ is $L_{adv}^G = L_{bce}(D(G(x)),1)$  where $x \sim p_{data}$ and $L_{bce}$ is the binary cross-entropy loss. The expression for $L_R$ and $L_{KL}$ are given in Equation \ref{lr} and Equation \ref{lkl} respectively. Therefore, the objective that needs to be minimized by the VAE part is $Loss = L_R + w_{KL}L_{KL} + L_{adv}^G$.
\begin{eqnarray}
L_{KL} & = & - \frac{1}{2N_z} \bigg(1 + \hat{\sigma} - \mu^2 - exp(\hat{\sigma})\bigg) \label{lkl} \\
L_R & = & L_s + L_p \quad \text{where, } \label{lr} \\
L_s & = & - \frac{1}{N_{max}} \sum_{i=1}^{N_s} log \big(  \mathcal{N}(\Delta x_i, \Delta y_i | \mu_{x,i}, \mu_{y,i}, \sigma_{x,i},\sigma_{y,i}) \big) \nonumber \\
L_p & = & - \frac{1}{N_{max}} \sum_{i=1}^{N_{max}} \sum_{k=1}^3 p_{k,i} log(q_{k,i}) \nonumber
\end{eqnarray}
where $w_{KL}$ is set to 0.5. On the other hand, the objective $L_{adv}^D$ that the discriminator minimizes is given as:
\begin{equation}
L_{adv}^D  =  L_{bce}(D(G(x)),0) + L_{bce}(D(x),1), \quad \text{where $x \sim p_{data}$.}
\end{equation}

The weights of the encoder and the decoder are updated using Adam whereas those of the discriminator are updated using SGD. It has been found out empirically by \cite{sketchrnn} that annealing $L_{KL}$ yields better results by making Adam optimizer to focus on minimizing the reconstruction loss term, which is tougher than optimizing KL Divergence loss. Therefore the modified objective is given as:
\begin{equation}
Loss  = L_R + w_{KL} \eta_{t} L_{KL} + L_{adv}^G, \quad \text{where} \myspace \eta_{t}  = 1 - (1 - \eta_{min})R^{t}
\end{equation}

The batch size is set to 100. $N_{max}$ is set to the length of the longest sequence in the dataset and is the maximum length of any generated sequence. Gradients are clipped between $[-1,1]$ for both generator and discriminator to avoid exploding of gradients, which is a common issue with sequence models. The initial learning rate is set to 0.001, with a decay of 0.9999 for every 100 iterations. The learning rate is decayed only if it is above a minimum threshold of 0.00001. The length of the latent vector $N_z$ is set to 128. In order to encourage the optimizer to put less focus on optimizing $L_{KL}$, it is modified as $ w_{KL} \eta_{t} max(L_{KL},KL_{min})$. The $KL_{min}$ is assigned a value of 0.2. Recurrent dropouts with drop probability of 0.1 is used.

\section{Experiments and Results}
\vspace{-3pt}
\subsection{Datasets, Baselines and Performance Metrics}
\vspace{-3pt}
We have used the QuickDraw Dataset created in \cite{sketchrnn} for training and experimentation. QuickDraw consists of sketches belonging to 345 different categories. Each category consists of 75000 sketches for training, 2500 for validation and 2500 for testing. As of today, QuickDraw is the only dataset with a large number of sketches in vector format for training and testing. We trained VASkeGAN and SkeGAN on the categories of sketches such as cat, firetruck, mosquito and yoga poses, as done in \cite{sketchrnn}. Since there is huge variety of sketches in QuickDraw, we chose these categories because they represent the sketches of humans, animals, insects and non-living things, and capture the diversity of the dataset. We have trained a separate model for each of the categories for both VASkeGAN and SkeGAN architectures, as done for Sketch-rnn in \cite{sketchrnn}. VASkeGAN was trained for 200000 iterations on the aforesaid sketch categories. The total number of training rounds for cat, mosquito, yoga and firetruck sketches are 4, 3, 6 and 4 respectively. The results of SkeGAN and VASkeGAN along their implications are discussed subsequently. We have quantitatively assessed the visual appeal of the sketches by performing a Human Turing Test with a group of 45 human subjects, to rate the sketches on a scale of 1 -- 5 on the categories such as clarity, drawing skill and naturalness. We have also introduced a metric, Ske-score, to objectively assess sketch generation in vector format.


\subsection{Results}
\begin{figure}[!htb]
\centerline{\includegraphics[scale=0.35]{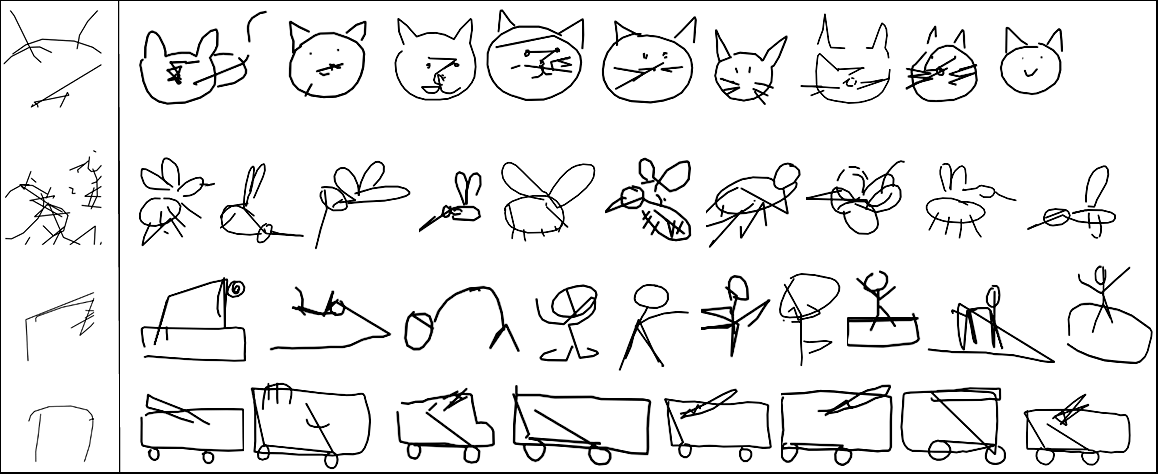}}
\caption{Sketches generated by SkeGAN after pre-training (left) and after the actual training (right).}
\label{seqganresults}
\end{figure}

\vspace{-3pt}
\noindent \textbf{Unconditional Generation of SkeGAN:}
All of the sketches are generated with a single starting tuple $S_0$. Subsequently tuples are generated until the pen-state $q_3$ equals 1 or the number of tuples generated becomes $N_{max}$. Figure \ref{seqganresults} shows some of the sketches generated for the aforesaid categories. Note that the sketches to the left of the separating line in Figure \ref{seqganresults} are generated just after the pre-training of the $G_\theta$. The sketches to the right of the separating line are generated by the trained model. The visual appeal of the generated images on the right favours the combination of policy gradients and adversarial loss for generating sketches. The images on the left of separating line indicates that pre-training is essential but not sufficient to generate good sketches. It very clear from Figure \ref{seqganresults} that the `scribble effect' of \cite{sketchrnn} is alleviated by SkeGAN.

\begin{figure}[!htb]
\centerline{\includegraphics[scale=0.35]{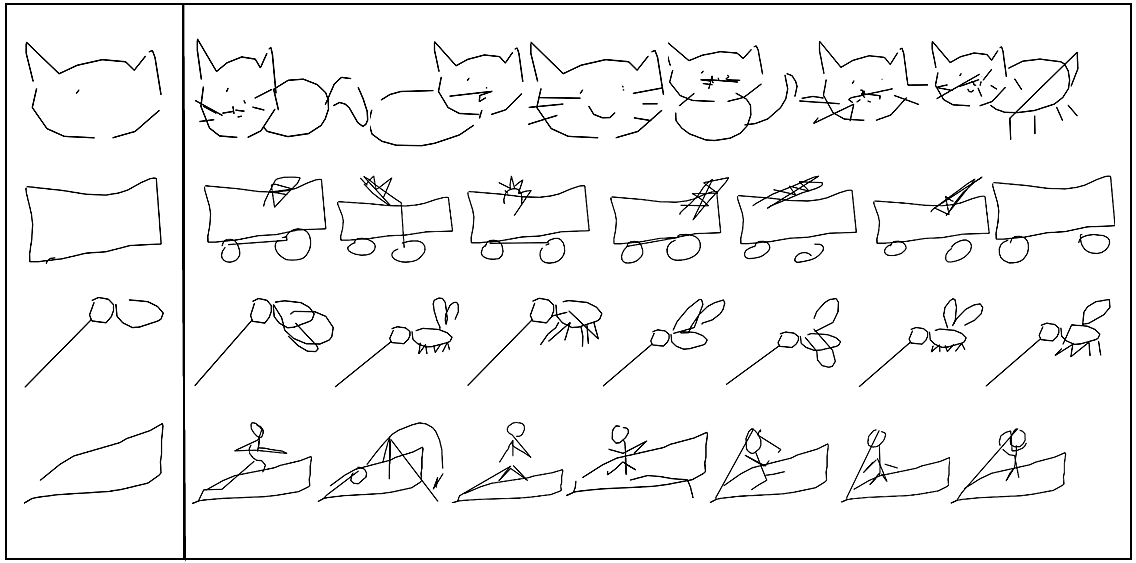}}
\caption{Partially drawn sketches (Left). Completed sketches by SkeGAN (Right).}
\label{sketchCompletion}
\end{figure}

\noindent \textbf{Sketch Completion by SkeGAN:}
In order to test the extrapolative abilities of SkeGAN, we feed a partially drawn sketch and observe how it can figure out various endings for the incomplete sketch. The generator which is trained with sketches of a particular category, is conditioned with an incomplete sketch from that category. The hidden state of the generator after this conditioning is $h$, which contains the semantic information of the incomplete sketch. Using this information, the remainder of the tuples for the sketch are sampled from the generator, with $h$ as its initial hidden state. Figure \ref{sketchCompletion} shows various completions for the same input sketch at a temperature $\tau$ of 0.25. The completed sketches shown are indeed meaningful and visually appealing, which highlights the creative aspect of SkeGAN.

\begin{figure}[!htb]
\centerline{\includegraphics[scale=0.4]{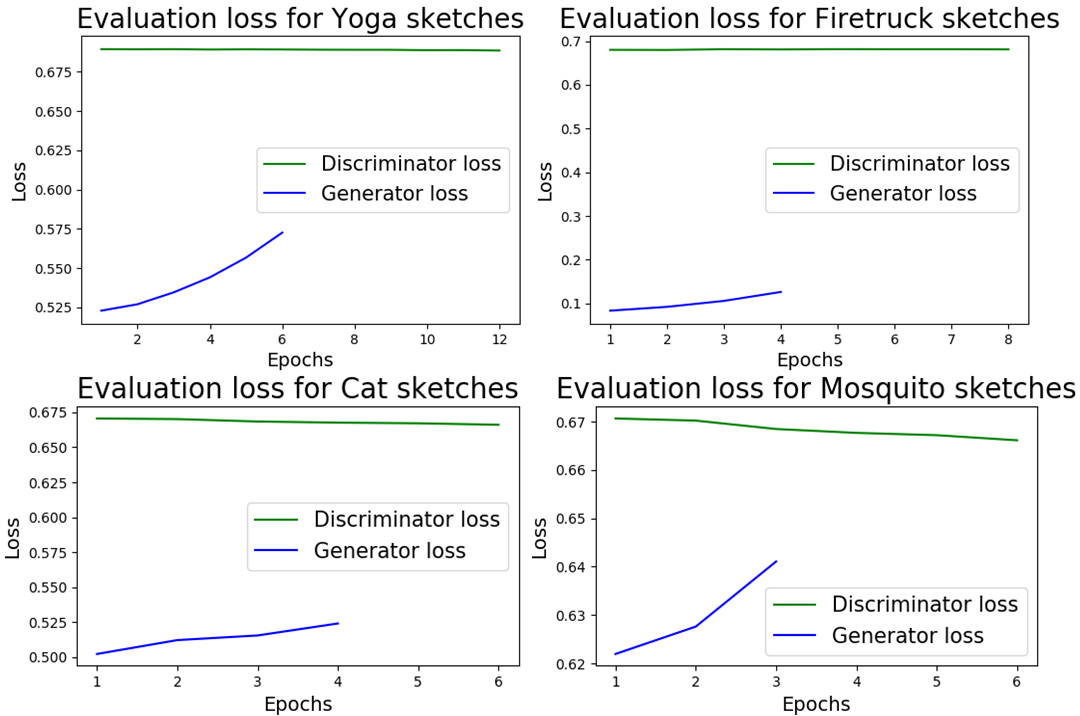}}
\caption{Evaluation loss of Generator and Discriminator for SkeGAN.}
\label{seqganeval}
\end{figure}

\noindent \textbf{Evaluation Loss of SkeGAN:}
Figure \ref{seqganeval} shows the evaluation loss for both the generator and the discriminator for across different categories of sketches. The generator is evaluated based on the reconstruction loss $L_R$ (as used in the pre-training stage), while the discriminator is evaluated using Negative Log Likelihood. The trends in both the evaluation losses point out to the fact that the minmax game played by the generator and discriminator is stable. Hence, the introduction of policy gradients into GAN has not affected the minmax game and the stability in training and has led to generation of visually appealing sketches.

\begin{figure}[!htb]
\centerline{\includegraphics[scale=.37]{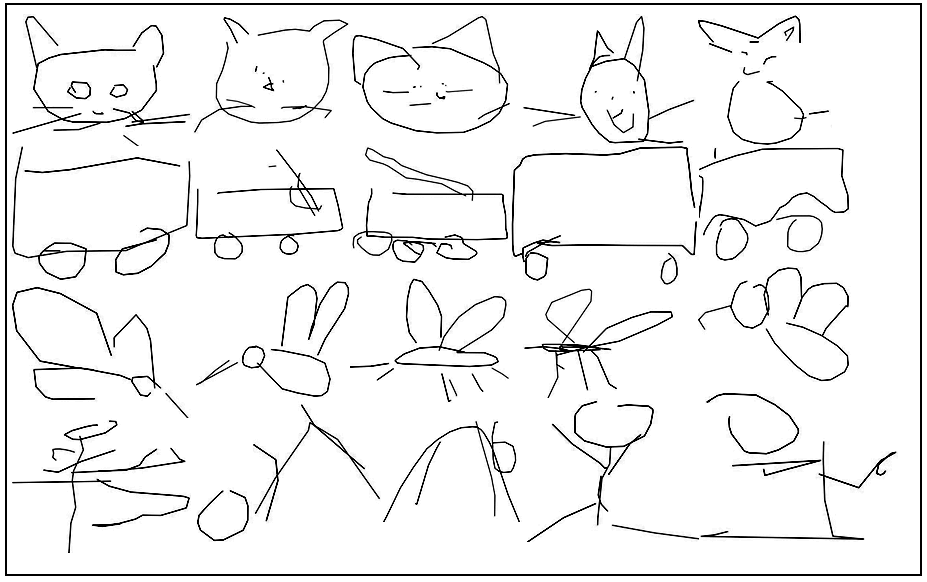}}
\centerline{\includegraphics[scale=.35]{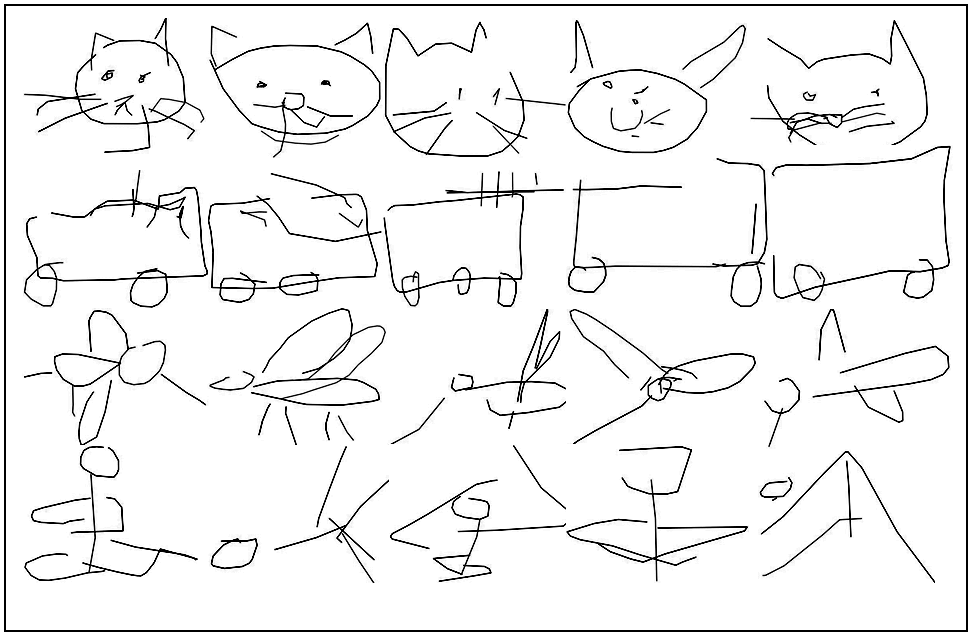}}
\caption{Sketches generated by VASkeGAN with GRU discriminator (Top) and LSTM discriminator (Bottom).}
\label{vaeganresults}
\end{figure}

\noindent \textbf{Sketches Generated by VASkeGAN:}
We allow the trained model to generate sketches after being conditioned by a sketch from a particular class. A sample of the generated images are shown in Figure \ref{vaeganresults}. This confirms that the model is indeed generating meaningful sketches and not random strokes. Here too, it is very clear from Figure \ref{vaeganresults} that, the `scribble effect' of \cite{sketchrnn} is alleviated by VASkeGAN. Following this, VASkeGAN is trained as a standalone GAN wherein the the encoder and the $L_{KL}$ are removed, retaining only the generator (decoder) and the discriminator with $L_R$ and adversarial loss. The sketches generated in this case for all the categories are just doodles without any discernible entity. This experimentally validates the fact that discrete outputs pose a difficulty for the gradient updates to be passed from discriminator to generator for the weight update. Hence empirically strengthening the argument in favour of the formulation of SkeGAN.

\begin{figure}[!htb]
\centerline{\includegraphics[scale=.35]{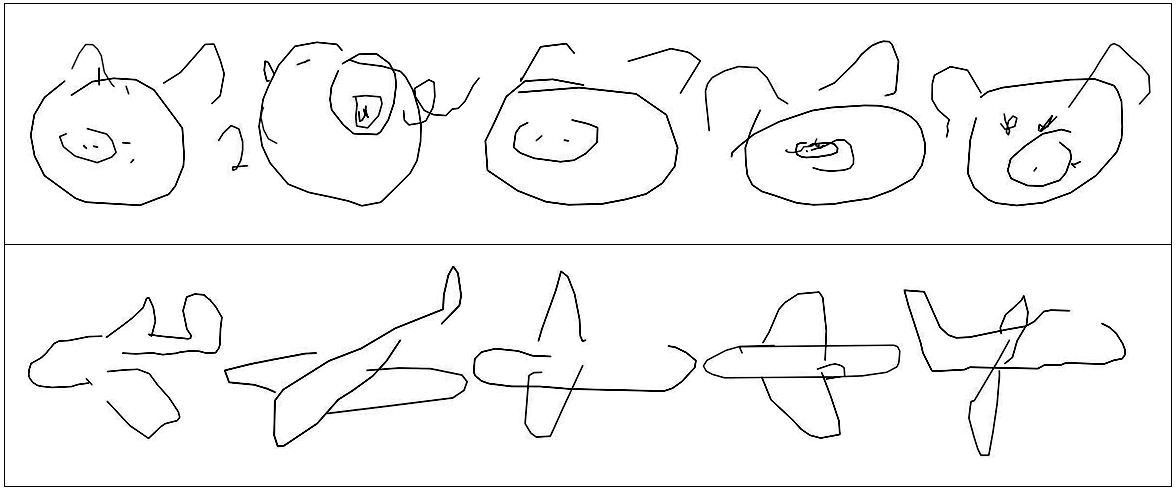}}
\centerline{\includegraphics[scale=.35]{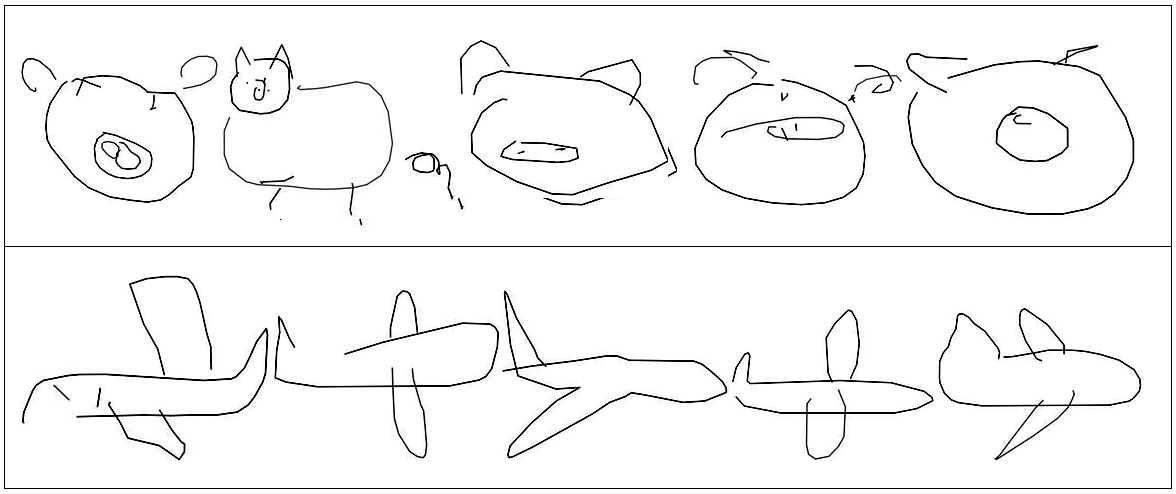}}
\caption{Transfer learning with GRU discriminator (Top) and LSTM discriminator (Bottom).}
\label{transfer}
\end{figure}

\noindent \textbf{Transfer Learning by VASkeGAN:}
The weights of the model trained on cat sketches for 200000 iterations, are used as an initialization to train two different models on pig and aeroplane sketches. In this case, the training is done for only 100000 iterations. Figure \ref{transfer} shows the pig and aeroplane sketches generated by transferring the learnt representations across categories. This shows that VASkeGAN generalizes well and is able to transfer the knowledge across categories. Transfer learning on SkeGAN led to a mode collapse, which is a direction of future investigation.

\begin{figure}[]
\centerline{\includegraphics[scale=0.4]{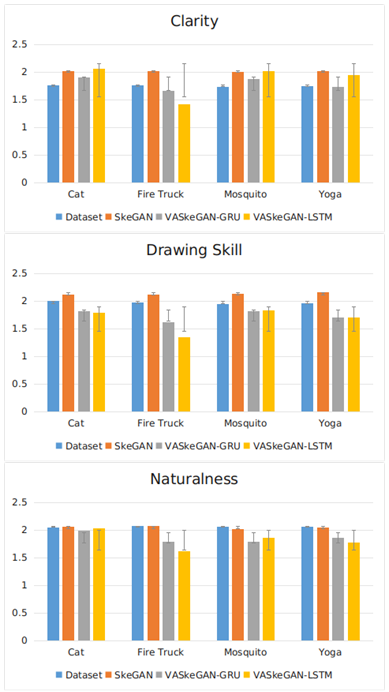}}
\caption{Scores of Human Turing Test for different criteria of visual appeal.}
\label{visualappealgraph}
\end{figure}

\noindent \textbf{Visual Appeal:} The scores of the Turing Test on the scale of 5 for the aforesaid criteria of visual appeal across the 4 categories of sketches are shown in Figure \ref{visualappealgraph}. The average of scores for a particular criterion across different categories are tabulated in Table \ref{turingtestvisualappeal}. It is interesting to note that SkeGAN generated sketches are almost at par with the those in the dataset. This supports our claim that SkeGAN can generate sketches that are clear, as artistic and as natural as those from the dataset.

\begin{table}[!htb]
\begin{center}
\begin{tabular}{|c|c|c|c|}
\hline
\textbf{Model} & \textbf{Clarity} & \textbf{Drawing Skill} & \textbf{Naturalness} \\ \hline
Dataset & 2.79 $\pm$ 1.01 & 2.71 $\pm$ 1.07  & 2.6 $\pm$ 1.12 \\ \hline
\textbf{SkeGAN} & \textbf{2.25 $\pm$ 1.1} & \textbf{2.25 $\pm$ 1.04} & \textbf{2.24 $\pm$ 1.21} \\ \hline
VASkeGAN(GRU) & 1.8 $\pm$ 0.86 & 1.74 $\pm$ 0.91 & 1.87 $\pm$ 1.12 \\ \hline
VASkeGAN(LSTM) & 1.87 $\pm$ 0.93 & 1.68 $\pm$ 0.79 & 1.83 $\pm$ 1.04 \\ \hline
\end{tabular}
\caption{Turing Test for Visual Appeal}
\end{center}
\label{turingtestvisualappeal}
\end{table}

\noindent \textbf{Evaluation of Sketches With Ske-score:}
The Ske-scores for the proposed models and for the dataset are tabulated in Table \ref{skescore}. It is clear that the Ske-scores of SkeGAN is closest to the dataset as compared to VASkeGAN. Therefore, SkeGAN generates `good' sketches without scribble effect. This shows that our formulation of SkeGAN is ideal for sketch generation in vector format.

\begin{table}[!htb]
\begin{center}
\begin{tabular}{|c|c|c|c|c|}
\hline
\textbf{Model} & \textbf{Cat} & \textbf{Fire truck} & \textbf{Mosquito} & \textbf{Yoga} \\ \hline
Dataset & 0.18 $\pm$ 0.07 & 0.12 $\pm$ 0.05 & 0.16 $\pm$ 0.08 & 0.15 $\pm$ 0.07 \\ \hline
\textbf{SkeGAN} & \textbf{0.19 $\pm$ 0.09} & \textbf{0.13 $\pm$ 0.08} & \textbf{0.18 $\pm$ 0.12} & \textbf{0.18 $\pm$ 0.1} \\ \hline
VASkeGAN (GRU) & 0.15 $\pm$ 0.07 & 0.1 $\pm$ 0.05 & 0.13 $\pm$ 0.08 & 0.12 $\pm$ 0.06 \\ \hline
VASkeGAN (LSTM) & 0.15 $\pm$ 0.06 & 0.09 $\pm$ 0.05 & 0.13 $\pm$ 0.06 & 0.13 $\pm$ 0.06 \\ \hline
\end{tabular}
\caption{Evaluation of generated sketches using Ske-score}
\end{center}
\label{skescore}
\end{table}

\section{Discussions}
\vspace{-4pt}
We now present ablation studies related to our models, discuss the best discriminator for VASkeGAN, and compare the training times of SkeGAN with VASkeGAN.\\

\begin{figure}[!htb]
\vspace{-3pt}
\centerline{\includegraphics[scale=.299]{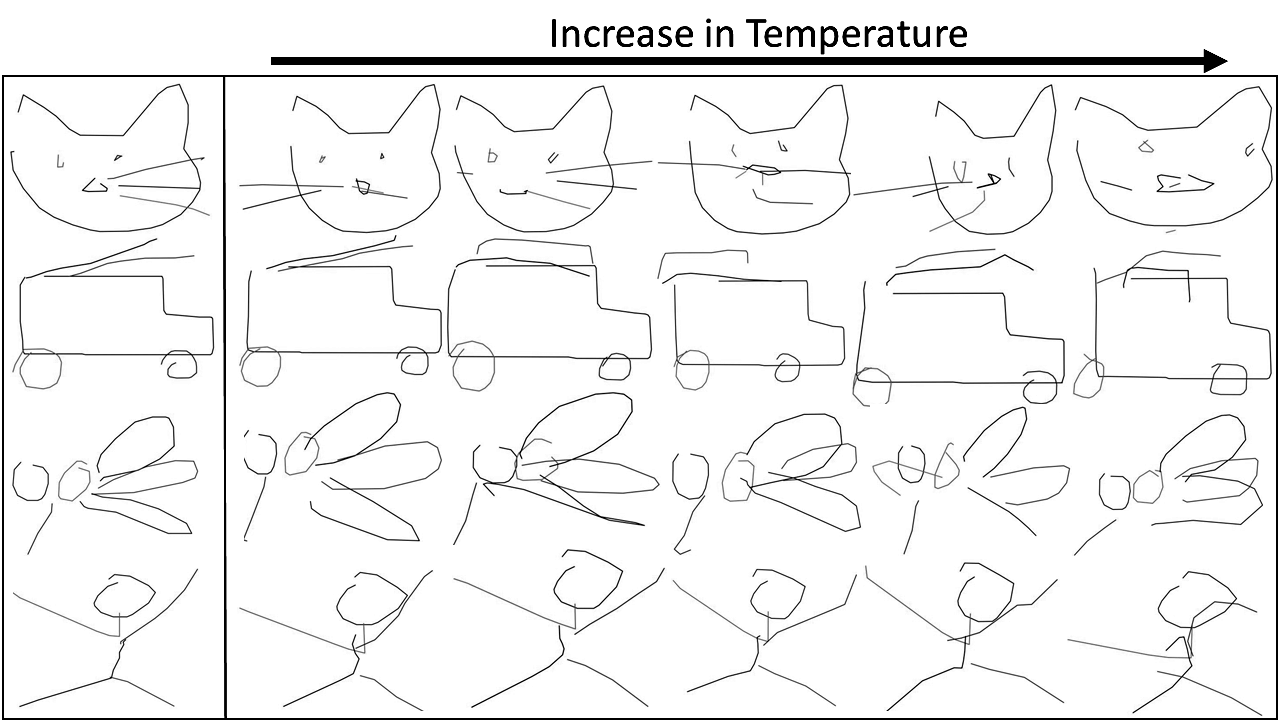}}
\centerline{\includegraphics[scale=.305]{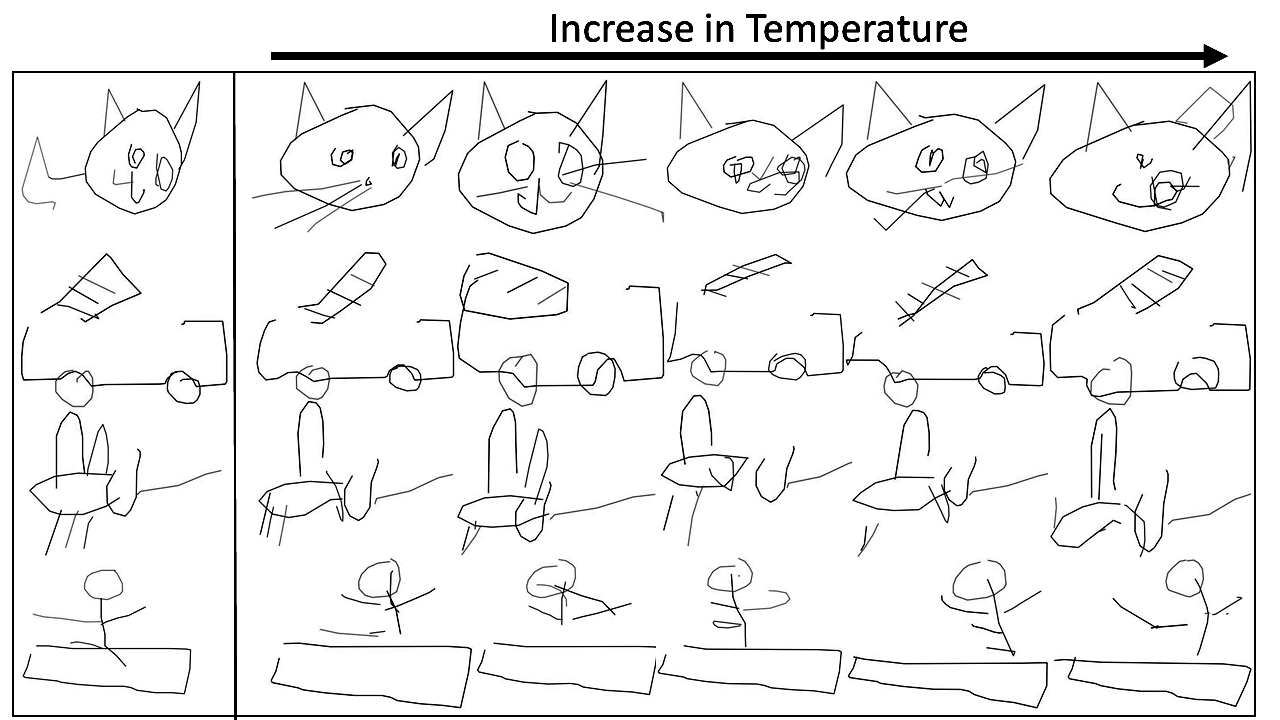}}
\caption{Effect of temperature on VASkeGAN with GRU discriminator (Top) and LSTM discriminator (Bottom).}
\label{vaegantempresults}
\end{figure}

\vspace{-5pt}
\noindent \textbf{Effect of Temperature $\tau$:}\\
\noindent \textit{Conditional Generation of VASkeGAN:}
We fix a sketch from each category and vary the temperature $\tau$ to see its effect on the reconstruction. The effect of temperature on sketches trained with GRU and LSTM discriminators is shown in Figure \ref{vaegantempresults}. The images to the left of the separating line in the top and bottom subfigures, are the human input to the trained model. To the right of the separating line in both the subfigures, are the ones that are conditionally generated by the VASkeGAN model at the temperatures of 0.2, 0.4, 0.6, 0.8 and 1.0 respectively. From the Figure \ref{vaegantempresults}, it can be noticed that as the temperature increases the ``randomness" increases. Under the influence of $\tau$, $\sigma^2_x$, $\sigma^2_y$ and $\hat{q}_k$ are replaced by $\sigma^2_x \tau$, $\sigma^2_y \tau$ and $\frac{\hat{q}_k}{\tau}$ respectively. Therefore, higher the value of $\tau$ more is the influence of variance which is translated into variations in the sketch generation.

Another observation is that the generated sketches of a particular category has the best visual appeal for a particular temperature. It is also at this temperature that along with reconstruction of sketches, extra visually appealing features (not present in the input image) are generated. For example, in Figure \ref{vaegantempresults}, the change in position of whiskers of cats in the top subfigure and the generation of whiskers on the cat's face in the bottom subfigure, are not present in the human input but are generated by reasoning out as to make the generated sketch more natural.

\begin{figure}[!htb]
\centerline{\includegraphics[scale=0.35]{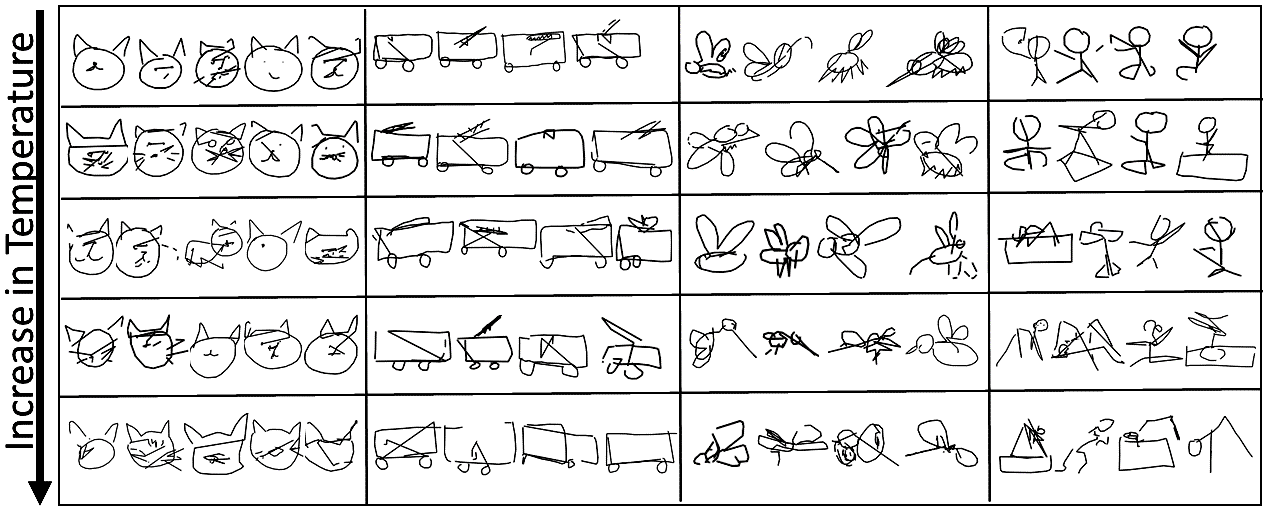}}
\caption{Effect of $\tau$ on sketches generated by SkeGAN.}
\label{seqgantempresults}
\end{figure}

\noindent \textit{Unconditional Generation of SkeGAN:}
The effect of temperature $\tau$ for unconditional generation is similar to its effect in conditional generation in the case of VASkeGAN. As $\tau$ increases the randomness of the generated sketches also increases. Since we are investigating its effect on unconditional generation, there is no ground-truth to compare the randomness. Instead, a group of sketches generated with a particular $\tau$ must be compared with those generated with a different value of $\tau$. The influence of $\tau$ on $\sigma_x$, $\sigma_y$ and $\hat{q}_k$ is same for those in conditional generation of VSkeGAN. In addition to this, it influences the mixing weights of GMM by acting as its inverse multiplier as in \cite{sketchrnn}. In Figure \ref{seqgantempresults}, there are 5 rows each depicting the sketches generated by SkeGAN at $\tau$ values of 0.2, 0.4, 0.6, 0.8 and 1.0. We find here that $\tau$ value of 0.4 is ideal for sketch generation based on visual appeal.

\begin{figure}[!htb]
\centerline{\includegraphics[scale=0.35]{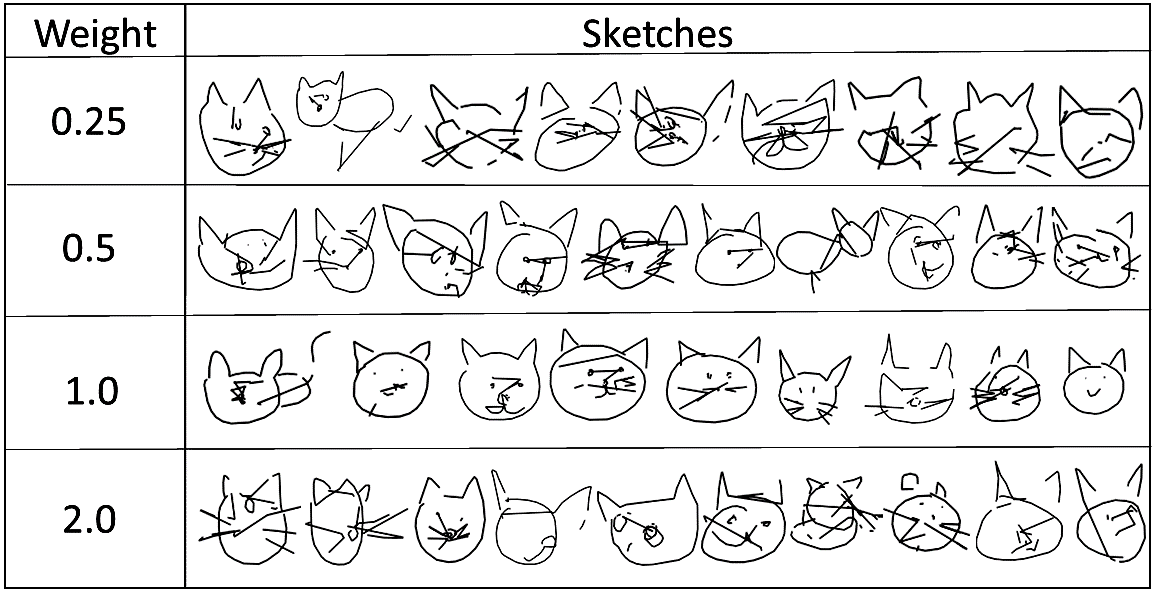}}
\centerline{\includegraphics[scale=0.35]{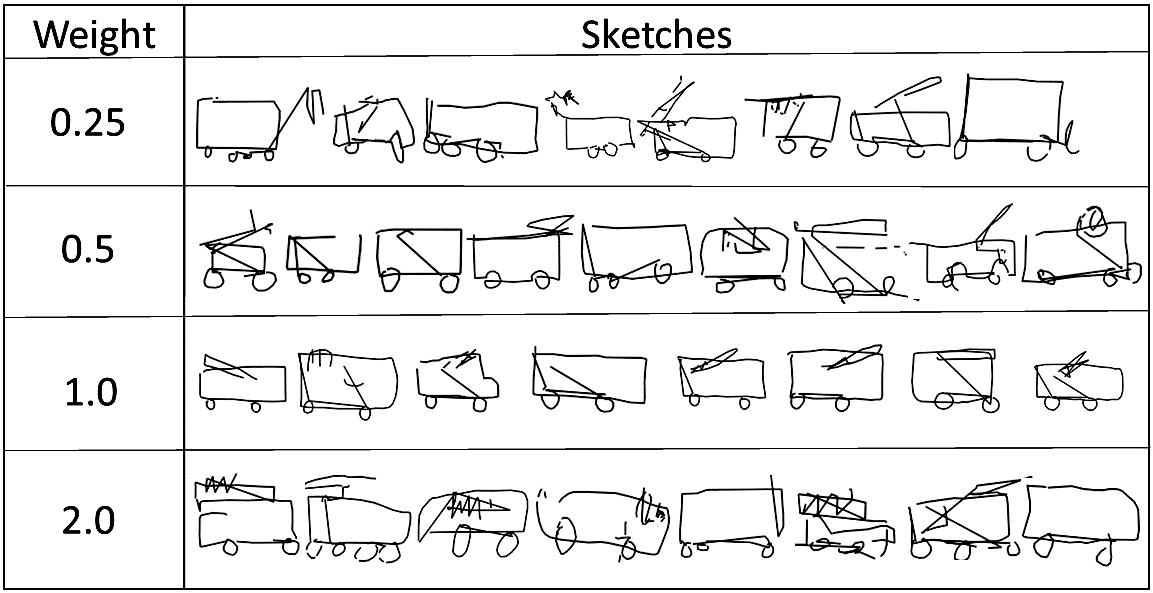}}
\caption{Effect of weighting policy gradient loss for cat sketches (Top) and firetruck sketches (Bottom).}
\label{effectWeightPGLossImg}
\end{figure}

\noindent \textbf{Weighting Policy Gradient Loss:} In order to understand the effect of policy gradient loss on the training, we multiply the loss with different weights and analyze its effect on the sketches generated and the Ske-score of the sketches. Figure \ref{effectWeightPGLossImg} shows the effect of multiplying weights to policy gradient loss on the sketches generated. One can observe that the weightage given to both policy gradient loss and adversarial loss must be equal in order to generate visually appealing sketches. Therefore, the ideal weight for both adversarial loss and policy gradient loss is 1.0 respectively.

The variation in Ske-scores due to variation in the weightage given to policy gradient loss in tabulated in Table \ref{effectWeightPGLossTab}. A very important observation is that as the weightage to the policy gradient loss is increased, the Ske-score increases. Also, a Ske-score which is sufficiently close to that of the dataset implies an alleviation of the Scribble Effect. Therefore, we conclude that the policy gradient formulation is ideal for pen-states as it reduces the Scribble Effect and helps in generating visually appealing sketches. The Scribble Effect is one of the short-comings of Sketch-rnn \cite{sketchrnn}.

\begin{table}[!htb]
\begin{center}
\begin{tabular}{|c|c|c|}
\hline
\textbf{Model} & \textbf{Cat} & \textbf{Firetruck} \\ \hline
Dataset        &   0.18 $\pm$ 0.07  & 0.12 $\pm$ 0.05  \\ \hline
SkeGAN@0.25    &    0.18 $\pm$ 0.08  & 0.13 $\pm$ 0.07  \\ \hline
SkeGAN@0.5     &    0.18 $\pm$ 0.08  & 0.13 $\pm$ 0.07   \\ \hline
SkeGAN@1.0     &    0.19 $\pm$ 0.09 &  0.13 $\pm$ 0.08  \\ \hline
SkeGAN@2.0     &    0.19 $\pm$ 0.09    & 0.13 $\pm$ 0.08   \\ \hline
\end{tabular}
\caption{Effect of weighting policy gradient loss on Ske-scores.}
\label{effectWeightPGLossTab}
\end{center}
\end{table}

\begin{figure}[!htb]
\centerline{\includegraphics[scale=.37]{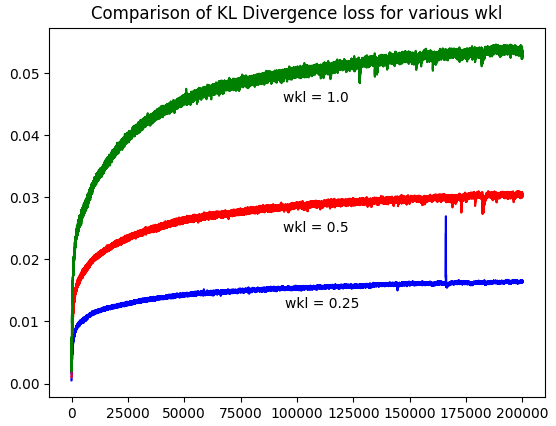}
\includegraphics[scale=.37]{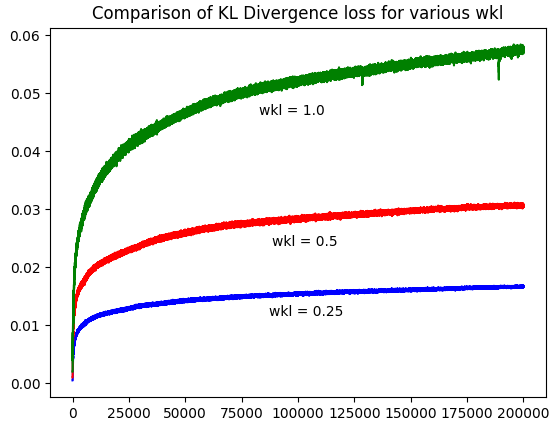}}
\caption{Plot of $L_{KL}$ for various $w_{KL}$ for GRU (left) and for LSTM (right) discriminators.}
\label{kldcat}
\end{figure}

\begin{figure}[!htb]
\centerline{\includegraphics[scale=.37]{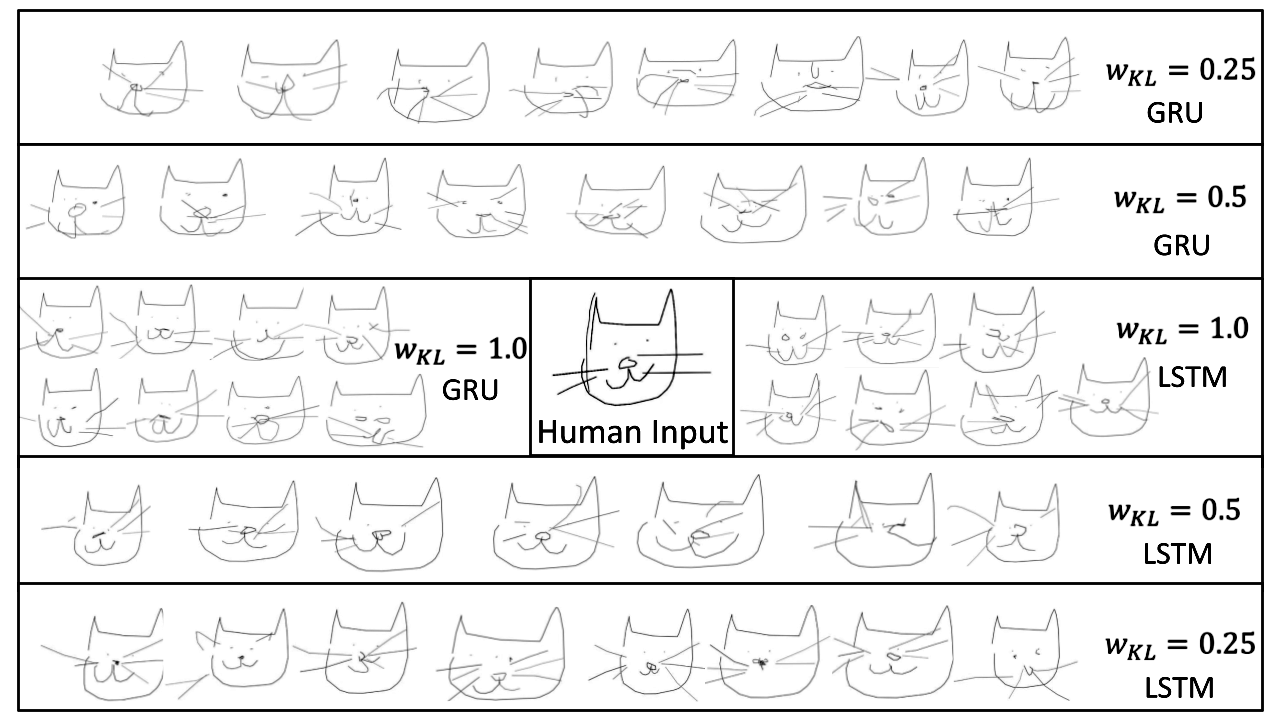}}
\caption{Effect of $w_{KL}$ on the generated sketches at constant $\tau$ of 0.25.}
\label{effectofwkl}
\end{figure}

\noindent \textbf{Weighting KL Divergence Loss:}
To understand the effect of weighting KL Divergence loss, we assign different values to $w_{KL}$ such as 0.25, 0.5 and 1.0 and analyze the quality of sketches generated both visually and using the Ske-score. Unlike \cite{sketchrnn}, where a higher $w_{KL}$ produces images closer to the data manifold, in our case, this behaviour is exactly opposite. Figure \ref{kldcat} shows the plot of $L_{KL}$ for different values of $w_{KL}$ while training the proposed model on cat sketches with GRU and LSTM respectively as the discriminators. The implication of this on the sketch generation is shown in Figure \ref{effectofwkl}. Visually inspecting these sketches, we conclude that a value of 0.5 is ideal for $w_{KL}$.

We also analyze the effect of changing $w_{KL}$ on the Ske-scores of the sketches. The values of Ske-scores of cat sketches for different $w_{KL}$ are tabulated in Table \ref{effectwklTab}. It can be observed that there is no significant change in Ske-score with change in $w_{KL}$ implying that there is no correlation between the weight assigned to $L_{KL}$ and the Ske-score.

\begin{table}[!htb]
\begin{center}
\begin{tabular}{|c|c|c|}
\hline
\textbf{Model} & \textbf{Cat} \\ \hline
Dataset & 0.18 $\pm$ 0.07 \\ \hline
VASkeGAN(GRU)@0.25 & 0.15 $\pm$ 0.06\\ \hline
VASkeGAN(GRU)@0.5 & 0.15 $\pm$ 0.07 \\ \hline
VASkeGAN(GRU)@1.0 & 0.15 $\pm$ 0.06  \\ \hline
VASkeGAN(LSTM)@0.25 & 0.15 $\pm$ 0.07 \\ \hline
VASkeGAN(LSTM)@0.5 & 0.15 $\pm$ 0.06\\ \hline
VASkeGAN(LSTM)@1.0 & 0.15 $\pm$ 0.06\\ \hline
\end{tabular}
\caption{Effect of $w_{KL}$ on Ske-Score.}
\label{effectwklTab}
\end{center}
\end{table}

\noindent \textbf{Discriminator for VASkeGAN:}
Based on the visual appeal scores for the sketches in each of the categories as shown in Figure \ref{visualappealgraph}, the best discriminator for VASkeGAN is an LSTM discriminator except for `firetruck' class where GRU worked well. This highlights the sensitivity involved in choosing a discriminator for training a GAN.\\

\vspace{-5pt}
\noindent \textbf{Training-time Comparison:}
Both VASkeGAN and SkeGAN are trained on \textit{NVIDIA GeForce GTX 1080 Ti}. The average time per iteration for training VASkeGAN is 0.6s, and hence a total time of 33.33 hours to train the model (for 200000 iterations) for a given category. The average time per iteration (one iteration of generator + two iterations of discriminator) for SkeGAN is 16.95s. The total time to train SkeGAN for cat, mosquito, yoga and firetruck sketches are 13.18 hours, 9.88 hours, 19.77 hours and 13.18 hours respectively, which are much lesser than the training time of VASkeGAN (33.33 hours). SkeGAN evidently admits faster convergence than VASkeGAN.

\section{Conclusion}
\vspace{-4pt}
In this work, we proposed two GAN-based approaches to address the problem of sketch generation in vector format. Until now, only a handful of approaches based on VAE \cite{ha2015recurrent,sketchrnn,chen2017sketch,zhong}, address this problem. We proposed two architectures viz. \textit{SkeGAN} and \textit{VASkeGAN}, the former a standalone GAN with policy gradients and adversarial loss, while the latter based on VAE-GAN. SkeGAN generates sketches that are better, both qualitatively and quantitatively, than those generated by VASkeGAN and has a faster convergence than VASkeGAN. SkeGAN generated sketches that are clear, natural and artistic compared to those from the dataset while maintaining a stable training process. Most importantly, both VASkeGAN and SkeGAN overcome the ``scribble effect" of \cite{sketchrnn}, thus highlighting their usefulness. Future directions include generalizing this work to larger vector-art datasets, including cartoons.

\bibliographystyle{unsrt}
\bibliography{references}

\begin{thebibliography}{10}

\bibitem{vae}
Diederik~P Kingma and Max Welling.
\newblock Auto-encoding variational bayes.
\newblock {\em arXiv preprint arXiv:1312.6114}, 2013.

\bibitem{gan}
Ian Goodfellow, Jean Pouget-Abadie, Mehdi Mirza, Bing Xu, David Warde-Farley,
  Sherjil Ozair, Aaron Courville, and Yoshua Bengio.
\newblock Generative adversarial nets.
\newblock In {\em Advances in neural information processing systems}, pages
  2672--2680, 2014.

\bibitem{draw}
Karol Gregor, Ivo Danihelka, Alex Graves, Danilo~Jimenez Rezende, and Daan
  Wierstra.
\newblock Draw: A recurrent neural network for image generation.
\newblock {\em arXiv preprint arXiv:1502.04623}, 2015.

\bibitem{dcgan}
Alec Radford, Luke Metz, and Soumith Chintala.
\newblock Unsupervised representation learning with deep convolutional
  generative adversarial networks.
\newblock {\em arXiv preprint arXiv:1511.06434}, 2015.

\bibitem{cyclegan}
Jun-Yan Zhu, Taesung Park, Phillip Isola, and Alexei~A Efros.
\newblock Unpaired image-to-image translation using cycle-consistent
  adversarial networkss.
\newblock In {\em Computer Vision (ICCV), 2017 IEEE International Conference
  on}, 2017.

\bibitem{yoo2016pixel}
Donggeun Yoo, Namil Kim, Sunggyun Park, Anthony~S Paek, and In~So Kweon.
\newblock Pixel-level domain transfer.
\newblock In {\em European Conference on Computer Vision}, pages 517--532.
  Springer, 2016.

\bibitem{ledig2017photo}
Christian Ledig, Lucas Theis, Ferenc Husz{\'a}r, Jose Caballero, Andrew
  Cunningham, Alejandro Acosta, Andrew Aitken, Alykhan Tejani, Johannes Totz,
  Zehan Wang, et~al.
\newblock Photo-realistic single image super-resolution using a generative
  adversarial network.
\newblock In {\em Proceedings of the IEEE Conference on Computer Vision and
  Pattern Recognition}, pages 4681--4690, 2017.

\bibitem{pix2pix}
Phillip Isola, Jun-Yan Zhu, Tinghui Zhou, and Alexei~A Efros.
\newblock Image-to-image translation with conditional adversarial networks.
\newblock {\em CVPR}, 2017.

\bibitem{pix2pixHD}
Ting-Chun Wang, Ming-Yu Liu, Jun-Yan Zhu, Andrew Tao, Jan Kautz, and Bryan
  Catanzaro.
\newblock High-resolution image synthesis and semantic manipulation with
  conditional gans.
\newblock In {\em Proceedings of the IEEE Conference on Computer Vision and
  Pattern Recognition}, 2018.

\bibitem{pathakCVPR16context}
Deepak Pathak, Philipp Kr\"ahenb\"uhl, Jeff Donahue, Trevor Darrell, and Alexei
  Efros.
\newblock Context encoders: Feature learning by inpainting.
\newblock In {\em Computer Vision and Pattern Recognition ({CVPR})}, 2016.

\bibitem{sketchrnn}
David Ha and Douglas Eck.
\newblock A neural representation of sketch drawings.
\newblock In {\em International Conference on Learning Representations}, 2018.

\bibitem{ha2015recurrent}
D~Ha.
\newblock Recurrent net dreams up fake chinese characters in vector format with
  tensorflow, 2015.

\bibitem{chen2017sketch}
Yajing Chen, Shikui Tu, Yuqi Yi, and Lei Xu.
\newblock Sketch-pix2seq: a model to generate sketches of multiple categories.
\newblock {\em arXiv preprint arXiv:1709.04121}, 2017.

\bibitem{zhong}
Kimberli Zhong.
\newblock {\em Learning to draw vector graphics: applying generative modeling
  to font glyphs}.
\newblock PhD thesis, Massachusetts Institute of Technology, 2018.

\bibitem{seqgan}
Lantao Yu, Weinan Zhang, Jun Wang, and Yong Yu.
\newblock Seqgan: Sequence generative adversarial nets with policy gradient.
\newblock In {\em Thirty-First AAAI Conference on Artificial Intelligence},
  2017.

\bibitem{vaegan}
Anders Boesen~Lindbo Larsen, S{\o}ren~Kaae S{\o}nderby, Hugo Larochelle, and
  Ole Winther.
\newblock Autoencoding beyond pixels using a learned similarity metric.
\newblock {\em arXiv preprint arXiv:1512.09300}, 2015.

\bibitem{sketchInterpretationRefinement}
Saul Simhon and Gregory Dudek.
\newblock Sketch interpretation and refinement using statistical models.
\newblock In {\em Rendering Techniques}, pages 23--32, 2004.

\bibitem{robotdrawing}
Patrick Tresset and Frederic~Fol Leymarie.
\newblock Portrait drawing by paul the robot.
\newblock {\em Computers \& Graphics}, 37(5):348--363, 2013.

\bibitem{sketchrecognition}
Ravi~Kiran Sarvadevabhatla, Jogendra Kundu, et~al.
\newblock Enabling my robot to play pictionary: Recurrent neural networks for
  sketch recognition.
\newblock In {\em Proceedings of the 24th ACM international conference on
  Multimedia}, pages 247--251. ACM, 2016.

\bibitem{sarvadevabhatla2016analyzing}
Ravi~Kiran Sarvadevabhatla et~al.
\newblock Analyzing structural characteristics of object category
  representations from their semantic-part distributions.
\newblock In {\em Proceedings of the 24th ACM international conference on
  Multimedia}, pages 97--101. ACM, 2016.

\bibitem{sarvadevabhatla2015eye}
Ravi~Kiran Sarvadevabhatla et~al.
\newblock Eye of the dragon: Exploring discriminatively minimalist sketch-based
  abstractions for object categories.
\newblock In {\em Proceedings of the 23rd ACM international conference on
  Multimedia}, pages 271--280. ACM, 2015.

\bibitem{eyefixation}
Ravi~Kiran Sarvadevabhatla, Sudharshan Suresh, and R~Venkatesh Babu.
\newblock Object category understanding via eye fixations on freehand sketches.
\newblock {\em IEEE Transactions on Image Processing}, 26(5):2508--2518, 2017.

\bibitem{pictionaryWord}
Ravi~Kiran Sarvadevabhatla, Shiv Surya, Trisha Mittal, and R~Venkatesh Babu.
\newblock Game of sketches: Deep recurrent models of pictionary-style word
  guessing.
\newblock In {\em Thirty-Second AAAI Conference on Artificial Intelligence},
  2018.

\bibitem{sketchparse}
Ravi~Kiran Sarvadevabhatla, Isht Dwivedi, Abhijat Biswas, Sahil Manocha, et~al.
\newblock Sketchparse: Towards rich descriptions for poorly drawn sketches
  using multi-task hierarchical deep networks.
\newblock In {\em Proceedings of the 25th ACM international conference on
  Multimedia}, pages 10--18. ACM, 2017.

\bibitem{gao2017ca-gan}
Jun Yu, Shengjie Shi, Fei Gao, Dacheng Tao, and Qingming Huang.
\newblock Composition-aided face photo-sketch synthesis.
\newblock 2017.

\bibitem{chen2018sketchygan}
Wengling Chen and James Hays.
\newblock Sketchygan: towards diverse and realistic sketch to image synthesis.
\newblock In {\em Proceedings of the IEEE Conference on Computer Vision and
  Pattern Recognition}, pages 9416--9425, 2018.

\bibitem{lu2018image}
Yongyi Lu, Shangzhe Wu, Yu-Wing Tai, and Chi-Keung Tang.
\newblock Image generation from sketch constraint using contextual gan.
\newblock In {\em Proceedings of the European Conference on Computer Vision
  (ECCV)}, pages 205--220, 2018.

\bibitem{quickdrawdataset}
Jonas Jongejan, Henry Rowley, Takashi Kawashima, Jongmin Kim, and Nick
  Fox-Gieg.
\newblock The quick, draw!-ai experiment, 2016.

\bibitem{sutton}
Richard~S Sutton, David~A McAllester, Satinder~P Singh, and Yishay Mansour.
\newblock Policy gradient methods for reinforcement learning with function
  approximation.
\newblock In {\em Advances in neural information processing systems}, pages
  1057--1063, 2000.

\bibitem{celeba}
Ziwei Liu, Ping Luo, Xiaogang Wang, and Xiaoou Tang.
\newblock Deep learning face attributes in the wild.
\newblock In {\em Proceedings of International Conference on Computer Vision
  (ICCV)}, 2015.

\bibitem{LFW}
Gary~B. Huang, Manu Ramesh, Tamara Berg, and Erik Learned-Miller.
\newblock Labeled faces in the wild: A database for studying face recognition
  in unconstrained environments.
\newblock Technical Report 07-49, University of Massachusetts, Amherst, October
  2007.

\end{thebibliography}

\end{document}